\documentclass[a4paper]{article}

\usepackage{icrc2013}
\usepackage[english]{babel}

\title{The Heliometer of Rio de Janeiro in Operation - 2013 Calibration}

\shorttitle{Heliometer of Rio de Janeiro Calibration}

\authors{
C. Sigismondi$^{1,2}$,
}

\afiliations{
$^1$ Observat\'orio Nacional/MCTI \\
$^2$ ICRANet/Sapienza Universit\`a di Roma and Ateneo Pontificio Regina Apostolorum \\
}

\email{sigismondi@icra.it}

\abstract{Here the preliminary results of the calibration of the heliometric angle of the Heliometer in Rio de Janeiro are presented. 
They have been made in 2013 with a reference surveying rod, equipped with two metal spheres acting as artificial star, when illuminated by the Sun,
and with the method of drift-scan timing.}

\keywords{Sun, solar diameter, heliometer, solar astrolabe, geomagnetic storm, heliometric angle, calibration.}

\begin{document}
\maketitle


\section{Introduction}
We are involved in the research of diameter variations and, as it is rather usual, in such metrological measurements
the knowledge of the absolute value is more complicate.
Only in the studies of stellar evolution the accurate value of the solar diameter has an utilization, 
in the hydrodynamical codes.
The absolute value of the solar diameter is defined by the inflexion points of the Limb Darkening Function.\cite{bib:Hill}
Recently this definition has been extended to the ephemerides-measurements, 
made with the timing of solar eclipses and planetary transits.\cite{bib:raponi}
The timing in astronomy is more accurate than imaging, and eclipses and planetary transits have been considered as the 
more accurate way to measure the solar diameter: e.g. the transits of Mercury of 2003 and 2006 have been used
to measure the solar diameter in SOHO/MDI 676.78 nm window.\cite{bib:emilio}

\section{Heliometric angle calibration}

To calibrate the heliometric angle $\theta$ we operated in two ways: one using imaging of a fixed target and the other by using the 
drfit-scan timing of the solar images over the CCD on the focal plane.
This calibration can give either the absolute value of the solar diameter and the confirmation that the heliometric angle 
remains fixed during the years. This last opportunity is of paramount importance 
for measurements which should be consistent along
several decades, in order to bring some astrophysical value.

\subsection{Reference rod at finite distance}

A wooden rod has been provided with two spheres of metal, and located 116 m far from the telescope, 
in a fixed position in the campus of the Astronomical Observatory.
The two spheres act as artificial stars during a sunny day, 
while when the weather is cloudy the image reflected by the spheres is much larger, corresponding to all the visible sky.
The telescope, observing without the solar filter, can aim at this rod, identifying the pointlike sources.
Their distance is measured on the same focal plane, to calibrate the scale therein.
The limit of this measurement is given by the local air turbulence, and it can be improved by the statistics as much as we need.

The first series of calibration has been realized in the month of February 2013
by using the wooden rod with two metal spheres, located at 116 meters from the telescope on the roof of the 
main building of the old Observatorio Nacional in the campus of the Observatory of Rio de Janeiro.
The image of the rod has been put on focus by using a two-pinholes mask.\cite{bib:Sigismondi2002}
The distance of the two spheres in pixels was different when using another mask, 
with a larger separation between the two pinholes, but this effect is a simple parallactic effect.

The modification of 1 cm in the distance between the two pinholes, with respect to a rod located 116 m far,
corresponds to an angle of $1/11600$ radians$\approx 20$ arcsec, and it is consistent with the variation of the pixel distance. 

Another proof that is not the effect of an imaging obtained along different Petzval surfaces\cite{bib:wiki} at dfferent offaxis
has been obtained during an ordinary observational session of the solar diameter: the images of the Sun drifting along different paths
with respect to the holes, yield constant measured diameter.
The measured diameter $D$ in pixels is related to the distance $d$ in pixels between the two images, 
produced by each half of a parabolic mirror by the formula $D=H-d$, where $H=\theta\times F$ and $F$ is the focal length.  
The pixel distance between the two spheres of the reference rod is of the same magnitude of the distance between the two 
images of the Sun. 

Finally the measurements made with the rod, are a reference for future checks with both the masks.
But they are critically dependent on the collocation of the masks with respect of the axis of the tube that, due to the
geometry of the mirror, is not axis-symmetrical. The annular heliometer\cite{bib:avila} will be axis-symmetrical.
Moreover the opening of the telescope to allow measurements of Earthly objects, then without the solar filter, exposes the
optics to the dust, and it is to avoid as much as possible.

\subsection{Drift-scan timing}

The measurement made with the drift-scan method is therefore more welcome.
The drift-scan is already the ordinary acquisition mode for heliometric images.
Usually 50 images are recorded without telescope following motion, in order to individuate 
the heliolatitude of the measured diameter with respect to the East-West drift.
The solar image moves on the focal plane at an angular speed depending only by the 
solar declination and the true solar day duration in that particular instant, all quantities known 
with high precision from ephemerides.
Therefore the solar image velocity can be used to calibrate the pixel scale of the CCD very accurately by acquiring 
200-250 images in oder to have the passage of the four limbs of the two solar images at the edges of the field of view or on each CCD columun.
The distance $d$ between the two images of the Sun is measured by the analysis program in pixels and it is related 
to the solar diameter
by the equation $D+d=F\times\theta$ where $\theta$ is the heliometric angle, and $F$ the focal length of the telescope. 
This equation is identical to $D+d=H$ if we consider that $F$ is also invariant.

The angular distance $\theta$=(D+d)/F can be measured by timing with drift scan, 
d is also measured directly by the heliometer.
F is constant within one part over $10^5$ because the instrument is made by carbon fiber and its longitudinal coefficient of thermal expansion at 300-350 K 
is $\lambda \le 10^{-6}/K$.\cite{bib:carbon}

The advantage of drift scan method is the timing:
the solar images drift on the focal plane and, with respect to a given reference on the CCDs,
even if there are optical distortions in the line of sight, the distortions act as a systematic error, 
which is the same for the same space direction.
In other words the timing is not affected by local optical distortions.

This kind of approach to measure the wedge angle has not been exploited with the Solar Disk Sextant SDS,
where have been used ten internal reflections within the heliometric wedge (the prismatic objective) of a laser beam 
in order to know the angle of this wedge to a 0.1 arcsec of accuracy ($1978.94\pm 0.1$ arcsec).\cite{bib:sofia13}
The detectors used in the SDS are seven linear CCDs of 100 pixel each, 
while in the Heliometer of Rio de Janeiro we can use all the CCD in VGA mode $640\times480$ pixels, with a scale of 1.168 arcsec/pixel.
The errorbar attributed to the method used to measure the wedge angle is 0.1 arcsec.
This errorbar can eventually be as larger as 0.2 or 0.3 arcsec.
What it is important is that this angle remained constant along the years.

At the heliometer with the drift-scan timing we obtained the following heliometric angle $1953.5\pm 1.4$ arcsec,
which is a preliminary value obtained by C. Sigismondi
with the measurement made on June 19, 2013 (three scans after local noon 12:45 PM) under very clear sky conditions.
The error associated is the one of two independent measurements of the diameter which resulted  $1888.2\pm 2.1$ arcsec,
the ephemerides reference with standard solar radius is 1888.85 arcsec for the same instant. 
The agreement is perfect. 
To reduce the error more measurements have to be done on the same series of images.

\subsection{Anomalous refraction and uncertainty on heliometric angle}

The uncertainty associated to the measurement of the heliometric angle 
is the dispersion of the three independent measurements
realized in that day.
It is well known that each single measurement obtained with drift-scan can differ from another one
because of the effect of the local atmospheric turbulence, in particular at frequencies below 0.01 Hz.\cite{bib:sigi3}

These fluctuations can produce a slow shift of the image during the transit which affects the final measurement of the diameter.\cite{bib:sigi4} 
also well verified at the Heliometer 
with a 100 s continuative observation made by C. Sigismondi on April 16, 2013 at 10 AM local time, 
the longest series of 500 images available up to now owing
to the memory limits of the present acquisition system.
These effects have been treated as anomalous refraction\cite{bib:corbard} instead of being considered as the low frequency part of the seeing
spectrum.\cite{bib:sigi3}

 \begin{figure}[t]
  \centering
  \includegraphics[width=0.5\textwidth]{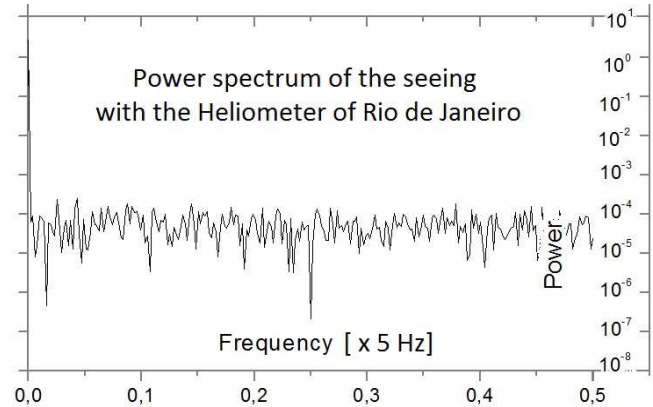}
  \caption{The power spectrum of the seeing as measured with the Heliometer of Rio de Janeiro. The unit of measure is the Nyquist frequence
  which is 5 Hz in our case. According to the Shannon theory half of this frequence if the limiting frequence at which we can get information.
  There is power at all frequencies, confirming the reliability of the hypothesis of low frequency motions, already verified with the Locarno telescope.
  \cite{bib:sigi4}}
  \label{simp_fig}
 \end{figure}

Considering these fluctuations as the low frequency region of the seeing seems a more logical approach, 
because it avoids to think to some strange effects of the atmosphere acting only somewhere, which are nothing else than 
phenomena ad hoc to explain the lack of coincidence between observations and expectations, too often invoked in past publications on solar astrometry.

The accuracy of 1.4 arcsec for this first measurements will allow to test the stability of the heliometric angle in the next months,
within this tolerance. A further complete analysis of the same data will allow to reduce the error 
of a statistical factor of $\sqrt500\sim 22$ reaching the desired 0.1 arcsec of accuracy.
The measurements made with the fixed rod on the top of the main builiding of the Museu de Astronomia will be the cross checks.

\section{Discussions and perspectives}

The Heliometer of Rio de Janeiro is already bringing new results to solar astrometry.
The quantitative discovery of glass filter effects has permitted to understand the shifts between the astrolabes of R2S3 network
and between them and SDS.
H. Neckel\cite{bib:neckel} showed that the variations of the solar diameter, in the continuum, 
does not exceed 0.07 arcsec within all the range of visibile wavelengths $\lambda$,
hence all departures larger than 0.1 arcsec remain unexplained. 
SODISM II experiment\cite{bib:corbard0613} in 2013 has confirmed the 0.07 arcsec range, but only after 
the measured diameters have been corrected for the diffraction (changing with $\lambda$) and
for the atmospheric turbulence (lower with increasing $\lambda$) 
acting over continuum limb darkening functions that are steeper for increasing $\lambda$:
the solar radii at 535.7 nm and at 607.1 nm are respectively $959.77 \pm 0.25$ and $959.83\pm 0.26$ arcsec. 
The reduction of the wavebands to a few nm in the PICARD/SODISM satellite
limitates the influence of emission lines from regions above the photosphere, 
but the differences of more than 0.5 arcsec within the data of various astrolabes
remained unexplained up to our verifications on the glass filter of the Heliometer of Rio de Janeiro.



The prediction of the space weather with an anticipation of a week for satellite in orbit around the Earth
 is a promising result and the reflecting Heliometer will provide in its observational duties. 

The study of low frequency component of the seeing is particularly suitable for the Heliometer configuration:
 defects of the tracking system or accidental hits or wind upon the tube act in the same  way for the two 
heliometric images of the Sun.
For single-image systems, there is always the doubt to observe a tracking defeat of the telescope.
Longer duration monitors of the seeing will clarify
the problem of consecutive meridian transits, with consecutive values separated often by more than 
the expected random errorbar determined by high frequency seeing. 

\vspace*{0.5cm}
\footnotesize{{\bf Acknowledgment:}{ C.S. acknowledges A. Raponi, the CNPq fund 300682/2012-3 and the Notre Dame Jerusalem Center.}}

\end{document}